\documentclass[aps,superscriptaddress,showpacs,preprint,amsmath,amssymb]{revtex4}
\usepackage{graphicx}
\usepackage{epsfig}
\usepackage{dcolumn}
\usepackage{bm}
\usepackage{subfigure}
\hfuzz=\maxdimen
\include{def-com}

\def \jp {J/\psi}

\def \gg {\gamma\gamma}

\def \ee {e^+e^-}
\def \uu {\mu^+\mu^-}

\def \ll {l^+l^-}
\def \mgev {\textrm{~GeV}/c^2}
\def \mmev {\textrm{~MeV}/c^2}
\newcommand{\chicj}{\chi_{cJ}}

\newcommand{\psip}{\psi^{\prime}}

\begin{document}
\title{\boldmath Precision measurements of branching fractions for $  \psip\to\pi^0\jp$ and $\eta\jp$}

\author{
\small{
M.~Ablikim$^{1}$, M.~N.~Achasov$^{5}$, O.~Albayrak$^{3}$, D.~J.~Ambrose$^{39}$, F.~F.~An$^{1}$, Q.~An$^{40}$, J.~Z.~Bai$^{1}$, Y.~Ban$^{27}$, J.~Becker$^{2}$, J.~V.~Bennett$^{17}$
% ...
\\
\vspace{0.2cm}
(BESIII Collaboration)\\
\vspace{0.2cm} {\it
$^{1}$ Institute of High Energy Physics, Beijing 100049, P. R. China\\
$^{2}$ Bochum Ruhr-University, 44780 Bochum, Germany\\
$^{3}$ Carnegie Mellon University, Pittsburgh, PA 15213, USA\\
% ...
$^{5}$ G.I. Budker Institute of Nuclear Physics SB RAS (BINP), Novosibirsk 630090, Russia\\
% ...
$^{17}$ Indiana University, Bloomington, Indiana 47405, USA\\
% ...
$^{27}$ Peking University, Beijing 100871, P. R. China\\
% ...
$^{39}$ University of Rochester, Rochester, New York 14627, USA\\
$^{40}$ University of Science and Technology of China, Hefei 230026, P. R. China\\
% ...
\vspace{0.4cm}
}
}
}

\vspace{2cm}
\begin{abstract}
We present a precision study of the $\psip\to\pi^0 J/\psi$
and $\eta J/\psi$ decay modes. The measurements are obtained using $106\times10^6$ $\psi'$ events accumulated with the BESIII detector at
the BEPCII $\ee$ collider operating at a center-of-mass energy corresponding to the $\psip$ mass.
We obtain $\mathcal{B}(\psip\to\pi^0 J/\psi)=(1.26\pm0.02{\rm~(stat.)}\pm0.03{\rm~(syst.)})\times 10^{-3}$ and
$\mathcal{B}(\psip\to\eta J/\psi)=(33.75\pm0.17{\rm~(stat.)}\pm0.86{\rm~(syst.)})\times 10^{-3}$.
The branching fraction ratio $R=\frac{\mathcal{B}(\psip\to\pi^0 J/\psi)}{\mathcal{B}(\psip\to\eta
J/\psi)}$ is determined to be $(3.74\pm0.06~{\rm(stat.)}\pm0.04~{\rm(syst.)})\times 10^{-2}$. The precision of these measurements of
$\mathcal{B}(\psip\to\pi^{0} J/\psi)$ and $R$ represent a significant improvement over previously published values.
\end{abstract}

\maketitle

\section{INTRODUCTION}
The study of the hadronic transitions between charmonium states has been an active field both for experimental and
theoretical research. The decays $\psip\to\eta\jp$ and $\pi^0\jp$ were first observed thirty years ago, and improved measurements of the corresponding branching fractions were performed by the BESII \cite{bes2} and CLEO \cite{cleo1}
collaborations. These decays are important probes of $\psip$ decay mechanisms that are characterized by the emission of a soft hadron. The QCD multipole-expansion (QCDME) technique was developed for applications to these heavy quarkonium system processes. For this, the measured branching
fraction for $\psip\to\eta\jp$ can be used to predict the $\eta$ transition rate between $\Upsilon$ states~\cite{kuangyp}.

The branching-fraction ratio, $R={\mathcal{B}(\psip\to\pi^0\jp)\over \mathcal{B}(\psip\to\eta\jp)}$, with $\mathcal{B}$ denoting the individual branching fraction, was
suggested as a reliable way to measure the light-quark mass ratio $m_u/m_d$~\cite{getqarkmass}.
Based on QCDME and the axial anomaly, the ratio is calculated to be $R=0.016$ with the conventionally accepted values of the quark masses
$m_s=150\mmev$, $m_d=7.5\mmev$ and $m_u=4.2\mmev$~\cite{miller}. Previously published measurements of this ratio give a significantly
larger value of $R=0.040\pm0.004$ \cite{pdg}. Recently, using chiral-perturbation
theory, the J\"ulich group investigated the source of charmed-meson loops in these decays as a possible explanation for this discrepancy~\cite{Feng-kunGuo:2009}.
Under the assumption that the charmed-meson loop mechanism saturates the $\psip\to\pi^{0}(\eta)J/\psi$ decay widths, they obtained
a value $R=0.11\pm0.06$, which indicates that the charmed-meson loop mechanism can play an important role in explaining the data. With parameters introduced into the
charmed-meson loop fixed using
$\mathcal{B}(\psip\to\eta\jp)$ as input, the hadron-loop contribution to
the isospin violation decay $\psip\to\pi^0\jp$ can be
evaluated~\cite{Guo:2012tj,Guo:2010ak}. Measurements of these
branching fractions can provide experimental evidence for hadron-loop contributions in charmonim decays, and impose more stringent constraints on charmed-meson loop contributions. It will also help clarify the influence of long-distance
effects in other charmonium decays, {\it e.g.}
$\psi(3770)\to\pi^0(\eta)\jp$~\cite{Zhang:2009kr,Guo:2012tj},~
$\psip\to\gamma\eta_c,$ and $~\jp\to\gamma\eta_c$~\cite{Li:2011ssa}.

This paper presents the most precise measurement of the ratio $R$ and the related branching fractions for $\psip\to\pi^0J/\psi$ and $\eta\jp$.

\section{BESIII EXPERIMENT AND DATA SET}
The BESIII experiment at the BEPCII \cite{NIM1} electron-positron collider is an upgrade of  BESII/BEPC \cite{besii}. The BESIII detector is designed to study hadron
spectroscopy and $\tau$-charm physics \cite{besphysics}. The
cylindrical BESIII spectrometer is composed of a Helium gas-based drift chamber
(MDC), a Time-Of-Flight (TOF) system, a CsI(Tl) Electromagnetic
Calorimeter (EMC) and a RPC-based muon identifier with a super-conducting
magnet that provides a 1.0 T magnetic field. The nominal detector acceptance is 93\% of
$4\pi$. The expected charged-particle momentum resolution and photon
energy resolution are 0.5\% and 2.5\% at 1 GeV, respectively. The
photon energy resolution at BESIII is much better than that
of BESII and comparable to that achieved by CLEO~\cite{cleo} and the
Crystal Ball~\cite{crysball}. An accurate measurement of photon
energies enables the BESIII experiment to study physics involving
photons, $\pi^0$ and $\eta$ mesons with high precision.

We use a data sample of (106.41$\pm$0.86)$\times$10$^6$ $\psip$ decays \cite{npsip}, corresponding to an integrated luminosity of
156.4 pb$^{-1}$. In addition, a 43 pb$^{-1}$ data sample collected at 3.65 GeV is used for QED background studies.

To optimize the event selection criteria and to estimate the
background, a {\sc geant4}-based simulation \cite{boost} is used that includes the geometries and
material of the BESIII detector components. An inclusive $\psip$ decay Monte Carlo (MC) sample is generated to study backgrounds. The generation of $\psip$ resonance production is simulated with
the MC event generator {\sc kkmc} \cite{kkmc}, while $\psip$
decays are generated with {\sc besevtgen} \cite{evtgen} for known decay
modes with branching fractions set to the
world average values \cite{pdg}, and with {\sc lundcharm} \cite{lundcharm} for the
remaining unknown decays. The analysis is performed in
the framework of the BESIII offline software system \cite{boss}
which handles the detector calibration, event reconstruction
and data storage.

\section{EVENT SELECTION}
Selection criteria described below are similar to those used in previous BES analyses \cite{chicj2vv,chicj2gv}. Candidate $\pi^0$ and $\eta$ mesons are reconstructed using two photons $\gg$, and the $\jp$ is reconstructed from lepton pairs $\ll$($l=e$ or $\mu$).

Photon candidates are reconstructed by clustering
EMC crystal energies. The energy deposited in nearby
TOF counters is included to improve the reconstruction
efficiency and the energy resolution. Showers identified as
photon candidates must satisfy fiducial and shower-quality
requirements. A minimum energy of 25~MeV is required for barrel showers ($|\cos\theta| < 0.8$) and 50 MeV for endcap
showers ($0.86 < | \cos\theta| < 0.92$). Showers in the angular range between the barrel and endcap
are poorly reconstructed and excluded from the analysis.
To exclude showers generated by charged particles, a photon
is required to be separated by at least $10^\circ$ from the nearest charged track.
EMC-cluster timing requirements are used to suppress
electronic noise and energy deposits unrelated to the event. The number of photons, $N_{\gamma}$, is required to be $N_{\gamma}\ge 2$.

Charged tracks are reconstructed from hit patterns in the MDC. The number of charged tracks is required to be two with zero net charge. For each track, the polar angle $\theta$ must satisfy
$|\cos\theta| < 0.93$, and the track is required to originate from within $\pm 10$~cm of the interaction point in the beam direction and within $\pm 1$~cm of the beam line in the plane perpendicular to the beam. The $J/\psi \rightarrow l^{+}l^{-}$ candidates are reconstructed from pairs of oppositely charged tracks.
Tracks are identified as muons (electrons) if their $E/p$ ratios satisfy $0.08~c< E/p < 0.22~c$ ($E/p > 0.8~c$), where $E$ and $p$ are the
deposited energy in the EMC and the momentum of the charged track, respectively.

To reduce the combinatorial background from uncorrelated $\gamma\gamma$ combinations and to improve the mass resolution, a four-constraint kinematic fit (4C-fit)
is applied with the hypothesis $\psip\to\gamma\gamma l^+l^-$ constrained to the sum of the initial $e^{+}e^{-}$ beam four-momentum. For events with more than two
photon candidates, the combination with the smallest $\chi^{2}$ is retained.

\begin{figure}[htbp]
\includegraphics[width=12cm,angle=0]{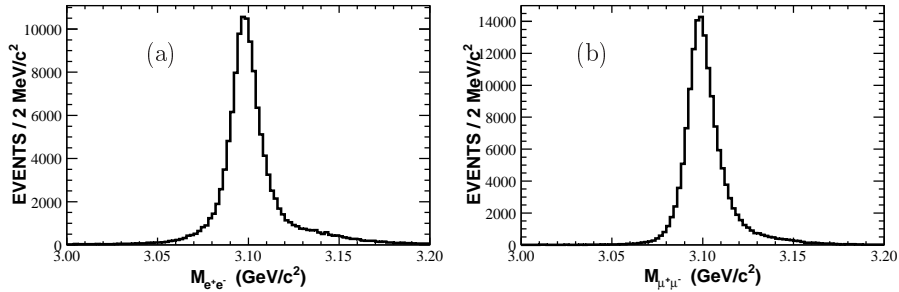}
\caption{The invariant-mass distributions for (a) electron-positron and (b) di-muon pairs in the selected $\gg\ll$ events in the data. \label{mll_pi0}}
\end{figure}

\begin{figure}[h]
\includegraphics[width=12cm,height=8cm]{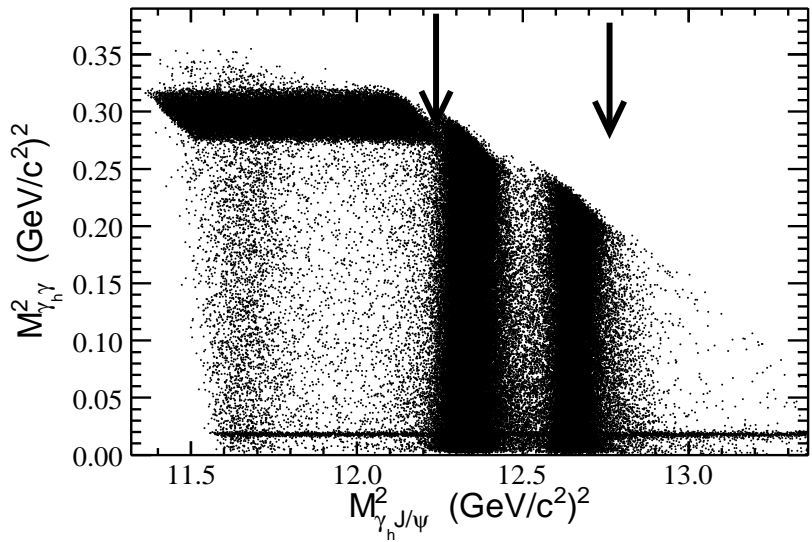}
\caption{\label{fig:scatter_plot}Dalitz plot of $M^2_{\gamma_h\gamma}$ (vertical) versus $M^2_{\gamma_h
\jp}$ (horizontal) for data, where $\gamma_h$ denotes the energetic photon. The horizontal bands around M$^{2}_{\gamma_{h}\gamma}$=0.02~(0.30) ~(GeV/c$^{2}$)$^{2}$ are due to $\psi' \rightarrow \pi^{0}(\eta) J/\psi$
 transitions. The vertical bands around $M^{2}_{\gamma_{h}J/\psi}$=11.65 (12.30, 12.70) (GeV/$c^{2}$)$^{2}$ are due to transitions $\psi' \rightarrow \gamma \chi_{c0(c1,c2)}$; the arrows denote the requirements to remove backgrounds from the $\chi_{c1,c2}$ states as described in the text.}
\end{figure}

The invariant-mass distribution for lepton pairs ($M_{ll}$) is shown in Fig.~\ref{mll_pi0}, where the $\jp$ signal is clearly seen with a
high signal to background ratio. For the further analysis, events are kept for which the reconstructed $\jp$ mass falls within a window of
$M_{ll}\in(3.05,~3.15)\mgev$; a mass window that is significantly larger than the mass resolution of about 8 MeV$/c^2$.
Figure~\ref{fig:scatter_plot} shows a Dalitz plot of the invariant-mass squared $M^{2}_{\gamma_h\jp}$ for the reconstructed $\jp$ and the energetic photon versus the two-photon invariant-mass squared $M^{2}_{\gamma_h\gamma}$, where $\gamma_h$ denotes the photon with the higher energy $E_{\gamma_h}>E_\gamma$.
Bands of $\pi^0,~\eta$, and $\chicj~(J=0,1,2)$ are clearly visible. To suppress the dominant source of background, which is from $\chi_{cJ}$ decays,
the mass of the $\gamma_h\jp$ system is required to satisfy the condition
$M_{\gamma_h J/\psi}\notin (3.50,~3.57)\mgev$ and $M_{\gamma_h\jp}<3.5\mgev$ for $\psip\to\pi^0\jp$ and $\eta\jp$, respectively. The 4C-fit $\chi^{2}$ is required to be less than $(100,~60,~70,~50)$
for the final states $(\pi^0\ee,~\pi^0\uu,~\eta\ee,~\eta\uu)$, respectively, where the values are determined by
optimizing the statistical significance $S/\sqrt{S+B}$, with $S(B)$ the number of signal (background) events. The background event levels are determined from the $\psip$ inclusive MC sample.

\section{Data analysis}
\label{sec:analysis}
\begin{figure}[hbtp]
\includegraphics[width=1\textwidth]{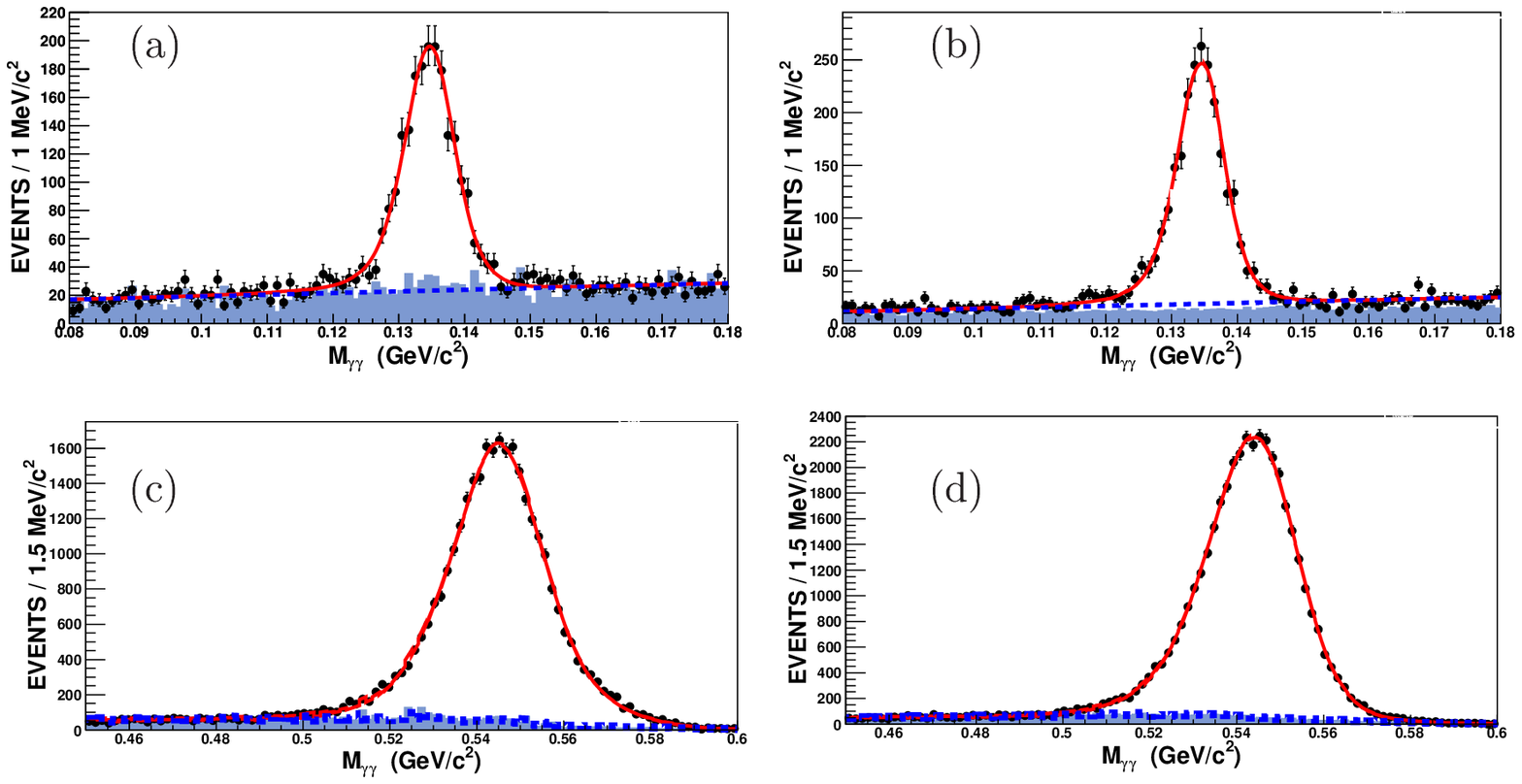}
\caption{(Color online) $M_{\gg}$ distributions and fit results. (a)~$\psip\to\pi^0 J/\psi,\jp\to \ee$, (b)~$\psip\to\pi^0 J/\psi,\jp\to \uu$,~(c)~$\psip\to\eta J/\psi,\jp\to \ee$,
(d)~$\psip\to\eta J/\psi,\jp\to \uu$, where the points with error bars are data, and the solid (red) curves are the total fit results, and the dashed curves are the fitted background shapes.
The hatched histograms represents dominant background events obtained from MC simulation and $\jp$ mass sidebands. \label{mggfit}}
\end{figure}

Background events from $\psip$ decays are studied using the inclusive MC sample. The background is
dominated by $\psip\to\gamma\chicj$, $\chicj\to\gamma\jp\to\gamma\ll$ decays. In addition, there are a few events from direct $\psip\to\gg\jp$, $\jp\to\ll$ decays \cite{ggjsi}. The shape of the $M_{\gamma\gamma}$ distribution of direct $\gamma\gamma\jp$ decays is smooth within both the $\pi^0$ and the $\eta$ mass regions.
The non-resonant background from $\psip\to\gg\ll$ is studied using $\jp$-mass sidebands in the data. For $\psip\to\eta\jp$,
there is an additional background from $\psip\to\pi^0\pi^0\jp$, which has a smooth shape within the $\eta$-mass
region. The background contribution from QED processes is studied using the continuum data taken at $\sqrt s=3.65 $ GeV, and it is found to be negligible.
The sum of all the MC-determined backgrounds in the $M_{\gamma\gamma}$ distribution are shown in Fig.~\ref{mggfit},
and they are found to be in reasonable agreement with those observed in the data.

\begin{figure}[hbtp]
\vspace{1cm}
\includegraphics[width=0.8\textwidth]{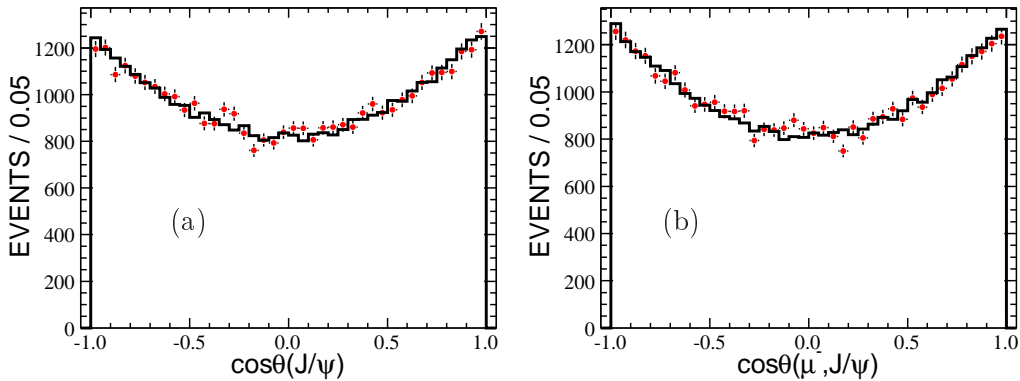}
\caption{(Color online) Angular distributions for (a) $\jp$ in the $\psip$ rest frame, (b) $\mu^-$ in the $\jp$ helicity system, where $\theta(\mu^-,\jp)$ is the angle between the $\mu$ momentum in $\jp$ rest frame and the $\jp$ momentum in $\psip$ rest frame. Points with error bars are data, and histograms are MC simulations as described in text.}
\label{angdis}
\end{figure}

To determine the detection efficiency, the angular distributions are properly modeled
in the event generator which accounts for polarization in the $\psip$ and $\jp$ decays. These decays are dominated
by their transverse polarization; longitudinal polarization of the $\psip$ is negligible due to its production from $\ee$ annihilation, and, since the $\jp$ is produced via $\psip\to\pi^0(\eta)\jp$ transitions,
its longitudinal polarization vanishes because of parity conservation. Thus, their polar-angle distributions take the form of
$dN/ d\cos\theta\propto (1+\cos^2\theta)$, where $\theta$ is the polar
angle of $\jp$ in the $\psip$ rest frame for $\psip\to\eta(\pi^0)\jp$ decays, or the angle between the lepton momentum in $\jp$ rest frame and the $\jp$ momentum in the $\psip$ rest frame for $\jp\to\ll$ decays. As an example, Fig.~\ref{angdis} shows angular distributions for $\jp$ and $\mu^-$ in $\psip\to\eta\jp\to\eta\mu^+\mu^-$ decays, where the angular distributions obtained from MC simulations (histograms) are observed to be in excellent agreement with the data (the points with error bars).
Similarly, we have verified that the angular distributions in the $\psip\to\pi^0\jp$ decay are well described by MC simulations. The detection efficiencies are determined using these MC event samples, and the values are listed in Table~\ref{fit_results}. The efficiencies for $\gg\ee$ final states are lower than that for $\gg\uu$ final states because the $e^+/e^-$ tracks suffer from stronger bremsstrahlung effects.

The signal yields are obtained from fits to the observed two-photon invariant mass M$_{\gamma\gamma}$ distributions. The observed line shapes are described with modified $\pi^0/\eta$ line shapes plus backgrounds. The $\pi^0$ and $\eta$ line shapes are taken from MC simulation including detector resolution; the $\pi^0$ and $\eta$ are described with relativistic Breit-Wigners in the event generation, and their masses and widths are fixed at their nominal values \cite{pdg}. To account for possible differences in mass resolution between data and MC simulation, the $\pi^0$/$\eta$ line shapes [LS$(\pi^0/\eta)$] are modified by convolution with a Gaussian function G($M_{\gamma\gamma}-\delta m,\sigma$). This technique of mass resolution smearing treatment was used in a previous publication~\cite{chicj2vv}. The probability distribution function (PDF) for the signal is taken as LS$(\pi^0/\eta)\otimes$G($M_{\gamma\gamma}-\delta m,\sigma$), where $\delta m$ and $\sigma$ correct the $\pi^0/\eta$ mass and mass resolution, respectively. The PDF for the dominant background contribution is obtained from MC
simulation and the residual background contribution is modeled as a first-order and a second-order polynomial function for the $\pi^0$ and $\eta$ channels, respectively.
The polynomial coefficients are free parameters with values determined from the data.
The fit results are shown in Fig.~\ref{mggfit}. For $\psip\to\pi^0 J/\psi$, the fit yields 1823$\pm$49 events for the $J/\psi\to\ee$ sample with a goodness of fit of $\chi^2/ndf=0.85$,
and 2268$\pm$55 events for $J/\psi\to\uu$ with a $\chi^2/ndf=0.86$, where $ndf$ denotes the number of degrees of freedom
in the fit. For $\psip\to\eta J/\psi$,
the fit yields 29598$\pm$202 events for $J/\psi\to\ee$ with a $\chi^2/ndf=1.33$ and 38572$\pm$280 events for $J/\psi\to\uu$
 with a $\chi^2/ndf=0.96$. The resulting values of $\delta m$ and $\sigma$ are $|\delta m|<1 \mmev$ and $\sigma<3\mmev$ for all
the modes. The signal yields are listed in Table \ref{fit_results}.

The branching fractions are calculated from the expression
\begin{equation}
\mathcal{B}(\psip\to X J/\psi) ={N^{sig} \over N_{\psip}\varepsilon \mathcal{B}(X\to\gamma\gamma) \mathcal{B}(J/\psi\to l^+l^-)},
\label{br_equation}
\end{equation}
where $X$ represents $\pi^0$ or $\eta$, $N^{sig}$ and $N_{\psip}$ are the signal yields and the number of $\psip$ events,
$N_{\psip}=106.41\times 10^6$. $\mathcal{B}(X\to\gamma\gamma)$ and $\mathcal{B}(J/\psi\to l^+l^-)$
denote the branching fractions of $\pi^0/\eta\to\gamma\gamma$ and $J/\psi\to\ee(\uu)$ \cite{pdg}. The variable $\varepsilon$
represents the MC-determined detection efficiency. The measured branching fractions for each final state are listed in Table \ref{ratio results}.

To validate the event selection criteria and fitting procedure, we perform a study using
a MC sample of 106$\times$10$^6$ inclusive $\psip$ events, with the known branching fractions as input.
The same analysis procedure as used for the real data is applied for this MC sample and the obtained branching fractions
for the $\psip\to\pi^0(\eta)\jp$ channels are found to be consistent with the input branching values within the
statistical accuracy.


\subsection{SYSTEMATIC ERRORS}
The main sources of systematic uncertainty originate from the number of $\psip$ events, the trigger efficiency,
the lepton tracking, photon reconstruction, kinematic fitting, uncertainties of the branching fractions for
$\pi^0(\eta)\to\gamma\gamma$ and $J/\psi\to\ee(\uu)$, and the selection and fitting procedures.

The uncertainty on the number of $\psip$ events is 0.81$\%$ as reported in Ref.~\cite{npsip}. Trigger efficiency uncertainty is 0.15$\%$ as reported in Ref.~\cite{trigger}.
The photon reconstruction uncertainty is determined to be 1\% per photon in Ref.~\cite{chicj2gv}, and, thus, the two-photon final state is assigned an uncertainty of 2$\%$. The tracking efficiency of the hard leptons is studied using a control sample of $\psip\to\pi^+\pi^-J/\psi$, $J/\psi\to\ee(\uu)$ decays. The tracking efficiency $\epsilon$ is calculated as $\epsilon = N_{full}/N_{all} \nonumber,$
where $N_{full}$ indicates the number of $\pi^{+}\pi^{-}l^{+}l^{-}$ events with all final tracks reconstructed successfully; and $N_{all}$ indicates
the number of events with one or both charged lepton tracks successfully reconstructed in addition to the pion-pair.
The difference in tracking efficiency between data and MC is calculated bin-by-bin over the distribution of transverse momentum versus the polar angle of the lepton tracks. By this method, tracking uncertainties are determined
to be 0.14$\%~$(0.20$\%$) and 0.16$\%~$(0.19$\%$) for $\psip\to\pi^0 J/\psi$,
$J/\psi\to\ee$($\uu$) and $\psip\to\eta J/\psi$, $J/\psi\to\ee$($\uu$), respectively.

Some differences are observed between data and MC $\chi^2$ distributions from the kinematic fit. These differences are mainly due to inconsistencies in the lepton track parameters between MC and data.
We apply correction factors for various $\mu^\pm~(e^\pm)$ track parameters that are obtained from control
 $\psip\to\pi^+\pi^-J/\psi$ data samples, where $J/\psi\to\ee(\uu)$. The correction factors are found by smearing the
MC simulation output so that the pull distributions properly describe those of the experimental data.
Half of the differences between the detection efficiencies, obtained using MC simulations with and without applying
these correction factors, are taken as systematic errors. These are 0.15$\%~$(0.19$\%$) and
0.20$\%~$(0.28$\%$) for $\psip\to\pi^0 J/\psi$,
$J/\psi\to\ee$($\uu$) and $\psip\to\eta J/\psi$, $J/\psi\to\ee$($\uu$), respectively.

Requirements on the $E/p$ ratio and the invariant mass $M_{l^+l^-}$ have been applied in the event selection. Uncertainties associated with these requirements are determined using the same control sample described above. Differences in the detection efficiency between the control data sample and the MC due to the $E/p$ ratio requirement are 0.06\% and 0.05\% for $J/\psi\to\ee$ and $J/\psi\to\uu$, respectively. Uncertainties caused by the mass window selection are
0.06\% for both the $\ee$ and $\uu$ channels.

An uncertainty due to the $M_{\gamma l^+l^-}$ requirement arises from a difference, $\delta(\chi_{c1,2})$, in the $\chi_{c1,2}$ mass resolution between the data and MC simulation, and is estimated by changing the MC-optimized requirement to one optimized using the data. Uncertainties caused by this $M_{\gamma l^+l^-}$ requirement are determined in this way to be 0.55\%~(0.16\%) and 0.11\%~(0.82\%)
for $\psip\to\pi^0 J/\psi$, $J/\psi\to\ee(\uu)$ and
$\psip\to\eta J/\psi$, $J/\psi\to\ee(\uu)$, respectively.

Systematic errors due to the background shape are estimated by varying the function used to describe nondominant backgrounds from a 1st- (2nd)- order polynomial to a 2nd- (3rd-) order polynomial for $\psip\to\pi^0(\eta)\jp$. The difference in the signal yields observed is taken as a systematic error. The uncertainty due to the choice of fitting range is estimated by repeating the fits using a fitting range that is 80\% as wide as that used in the original fit. The difference in the signal yields is taken as a systematic error.
Table~\ref{sys tot err} summarizes all the sources of systematic uncertainties.

\subsection{RESULTS AND DISCUSSION}
Branching fractions for the decays $\psip\to\pi^0 J/\psi$ and $\eta J/\psi$ with $\jp\to\ee,~\uu$ are calculated with Eq.~(\ref{br_equation}) using the fitting results and the detection efficiencies as inputs. Branching fractions measured using $\jp\to\ee$ and $\uu$ final states are combined together with the weighted average method described
in Ref.~\cite{pdg}, here common systematic uncertainties are counted only once.
The combined branching fractions are $\mathcal{B}(\psip\to\pi^0\jp)=(1.26\pm0.02\pm0.03)\times 10^{-3}$ and $\mathcal{B}(\psip\to\eta\jp)=(33.75\pm0.17\pm0.86)\times10^{-3}$ (see Table \ref{ratio results}). Using the measured branching fractions, the ratio $R=\mathcal{B}(\psip\to\pi^0 J\psi)/ \mathcal{B}(\psip\to\eta J\psi)$ is calculated to be $R=(3.74\pm0.06\pm0.04)\times10^{-2}$ (see
Table \ref{ratio results}). Note that systematic uncertainties that are common to both channels cancel in the ratio.
Our combined result on the $R$-ratio is consistent with previous world average values with a precision improvement of about a factor of five.

These precise measurements of the
$\psip\to\pi^0\jp$ and $\eta\jp$ branching fractions permit the study of isospin violation
mechanisms in the $\psip\to\pi^0\jp$ transition.
As shown in  \cite{Feng-kunGuo:2009,gfk_zq}, the axial anomaly does
not adequately explain the observed isospin violation, while
contributions from charmed meson loops would be a possible
mechanism for additional isospin violation sources. Confirmation of
sizeable contributions from charmed-meson loops would be an
indication that non-perturbative effects play an important
role in the charmonium energy region.

\begin{table}[htbp]
\setlength{\tabcolsep}{0.5pc}
\caption{Summary of signal yields and detection efficiencies for the each final state.\label{fit_results}}
\begin{tabular}{lllll}
 \hline\hline
Mode &\multicolumn{2}{c}{$\psip\to\pi^0
J/\psi$}&\multicolumn{2}{c}{$\psip\to\eta J/\psi$}\\\hline
Final state
&$\gamma\gamma\ee$&$\gamma\gamma\uu$&$\gamma\gamma\ee$&$\gamma\gamma\uu$\\\hline
$\varepsilon$(\%)&23.05&29.11&35.41&46.28\\
$N^{sig}$
&1823$\pm$49&2268$\pm$55&29598$\pm$202&38572$\pm$280\\\hline\hline
\end{tabular}
\end{table}

\begin{table}[htbp]
\setlength{\tabcolsep}{0.5pc}
\caption{Summary of measured branching fractions ($\mathcal{B}$) and the ratio $R={\mathcal{B}(\psip\to\pi^0 J\psi)\over
\mathcal{B}(\psip\to\eta J\psi)}$ with comparison to world average values (see PDG).\label{ratio results}}
\begin{center}
\begin{tabular}{lcccc}
\hline\hline
$\mathcal{B}$ or $R$ & Final state & This work & Combined & PDG\cite{pdg}\\\hline
$\mathcal{B}(\psip\to\pi^0\jp)$ & $\gg\ee$ & $1.27\pm0.03\pm0.03$&---&--- \\
$(\times10^{-3})$               & $\gg\uu$ & $1.25\pm0.03\pm0.03$& $1.26\pm0.02\pm0.03$&$1.30\pm0.10$ \\\hline
$\mathcal{B}(\psip\to\eta\jp)$ & $\gg\ee$ &$33.77\pm0.23\pm0.93$ &---&--- \\
$(\times10^{-3})$              & $\gg\uu$ &$33.73\pm0.24\pm0.90$ & $33.75\pm0.17\pm0.86$&$32.8\pm0.7$ \\\hline
$R={\mathcal{B}(\psip\to\pi^0\jp)\over \mathcal{B}(\psip\to\eta\jp)}$&$\gg\ee$&$3.76\pm0.09\pm0.06$&--&---\\
$(\times10^{-2})$ &$\gg\uu$&$3.71\pm0.09\pm0.05$&$3.74\pm0.06\pm0.04$&$3.96\pm0.42$\\\hline\hline
\end{tabular}
\end{center}
\end{table}

\begin{table}[htbp]
\caption{Summary of all systematic errors (\%) considered in this analysis. \label{sys tot err}}
\begin{center}
\begin{tabular}{lcccc}
 \hline\hline
 Sources&$\pi^{0} J/\psi(\ee)$&$\pi^{0}
 J/\psi(\uu)$&$\eta J/\psi(\ee)$&$\eta
 J/\psi(\uu)$\\\hline
 $N_{\psip}$&0.81&0.81&0.81&0.81\\
 Trigger&0.15&0.15&0.15&0.15\\
 Tracking&0.14&0.20&0.16&0.19\\
 Photon&2.00&2.00&2.00&2.00\\
 4-C Fit&0.15&0.19&0.20&0.28\\
 $B_{r}(J/\psi\to l^{+}l^{-})$&1.01&1.01&1.01&1.01\\
 $B_{r}(\pi^0/\eta\to\gamma\gamma)$&0.03&0.03&0.51&0.51\\
 M($l^+l^-$) &0.06&0.06&0.06&0.06\\
 M($\gamma l^+l^-$)&0.55&0.16&0.11&0.82\\
 E/p&0.06&0.05&0.06&0.05\\
 Background shape&0.24&0.24&1.14&0.10\\
 Fitting range&0.63&0.80&0.55&0.58\\ \hline
 Total&2.55 &2.55 &2.77 &2.66 \\\hline\hline
\end{tabular}
\end{center}
\end{table}

\newpage
{\bf Acknowledgement:}\\
The BESIII collaboration thanks the staff of BEPCII and the computing center for their hard efforts. This work is supported in part by the Ministry of Science and Technology of China under Contract No. 2009CB825200; National Natural Science Foundation of China (NSFC) under Contracts Nos. 10625524, 10821063, 10825524, 10835001, 10935007, 11125525; Joint Funds of the National Natural Science Foundation of China under Contracts Nos. 11079008, 11179007; the Chinese Academy of Sciences (CAS) Large-Scale Scientific Facility Program; CAS under Contracts Nos. KJCX2-YW-N29, KJCX2-YW-N45; 100 Talents Program of CAS; Istituto Nazionale di Fisica Nucleare, Italy; Ministry of Development of Turkey under Contract No. DPT2006K-120470; U. S. Department of Energy under Contracts Nos. DE-FG02-04ER41291, DE-FG02-91ER40682, DE-FG02-94ER40823; U.S. National Science Foundation; University of Groningen (RuG) and the Helmholtzzentrum fuer Schwerionenforschung GmbH (GSI), Darmstadt; WCU Program of National Research Foundation of Korea under Contract No. R32-2008-000-10155-0

\end{document}